\documentclass[twocolumn,groupedaddress,
amsmath,amssymb,prl]{revtex4-2}
\usepackage{graphicx}%
\usepackage[pdftex,colorlinks=true,linkcolor=blue,citecolor=blue,urlcolor=blue]{hyperref}
\begin{document}
\title{ Topological bound states in the continuum with controllable multiplicity}
\author{Ya-Ping Lou}
\author{Wei Jia}
\email{jiaw@lzu.edu.cn}
\affiliation{Lanzhou Center for Theoretical Physics, Key Laboratory of Theoretical Physics of Gansu Province, Key Laboratory of Quantum Theory and Applications of MoE, Gansu Provincial Research Center for Basic Disciplines of Quantum Physics, Lanzhou University, Lanzhou 730000, China}

\begin{abstract}

Bound states in the continuum (BICs) are spatially localized states embedded in the continuous spectrum without hybridizing with extended bulk modes. Recent advances in topological band theory have greatly enriched the understanding of BICs, which gives rise to boundary-localized topological BICs with extremely high robustness against disorders. However, there remains a challenge in realizing corner-localized topological BICs in a three-dimensional system due to the absence of both realistic theoretical models and effective topological characterization schemes. In particular, how to engineer a controllable number of corner-localized topological BIC is still an open question. Here, we propose that the corner-localized topological BICs can emerge in a class of generalized breathing pyrochlore lattice with general inter-cell hoppings. We further show that the number of BICs at each corner can be arbitrarily adjusted by changing the parameters of inter-cell hoppings. Remarkably, although these corner-localized topological BICs are intertwined with a substantial number of bulk modes, we can accurately characterize them through the polarized topological charges, which are nodal points with topological properties in Brillouin zone and are measurable in experiments. We also reveal three types of topological phase transitions of corner-localized BICs, which are associated with the different ways of closing the bulk energy gap and can be intuitively captured by the polarized topological charges. This work not only promotes the theoretical research of corner-localized topological BICs, but also opens an avenue for their experimental observation in the future.

\end{abstract}
%\date{\today}

\maketitle

{\it\color{blue}Introduction.}~Nearly one century ago, the concept of bound states in the continuum (BICs) was first proposed by von Neumann and Wigner, who demonstrated that an engineered quantum potential can support spatially localized states embedded in a continuous spectrum~\cite{von1929}. Unlike ordinary resonant states that radiate into the continuum, BICs remain perfectly localized despite coexisting with extended propagating modes at the same energy~\cite{PhysRevA.31.3964,PhysRevA.32.3231,hsu2016bound}. Such a counterintuitive phenomenon has stimulated the sustained interest across a broad range of physical platforms, including photonic~\cite{PhysRevLett.100.183902,hsu2013observation,PhysRevLett.112.213903,PhysRevLett.121.253901,PhysRevLett.122.187402,yu2020acoustoopic}, acoustic~\cite{PhysRevB.97.024304,huang2021sound,PhysRevB.106.085404,PhysRevApplied.18.054021,PhysRevApplied.19.054001,martisabate2024observation,PhysRevB.110.054201}, electronic~\cite{PhysRevA.73.022113,PhysRevB.83.235321,PhysRevB.85.115307,PhysRevB.92.245107,li2020bound,PhysRevB.110.035428,PhysRevB.111.L060101}, and wave systems~\cite{linton2007embedded,PhysRevLett.132.187202}. Recently, a large number of studies have revealed that BICs originate from two fundamental mechanisms, i.e., symmetry-protected BICs~\cite{PhysRevLett.107.183901,PhysRevLett.109.067401} and parameter-tuned BICs~\cite{dreisow2009adiabatic,zhao2020terahertz}, inducing a classification based on the differences in their decoupling ways. Moreover, the rapid development of artificial lattices and metamaterials has established BICs as a universal wave phenomenon with promising applications in high-$Q$ resonators~\cite{PhysRevLett.119.243901,PhysRevLett.121.193903,jin2019topologically,PhysRevLett.126.117402,fan2025active}, low-threshold lasing~\cite{zhang2020lowthreshold,hwang2021ultralow,ren2022lowthreshold}, nonlinear enhancement~\cite{PhysRevLett.121.033903,koshelev2020subwavelength}, and robust energy confinement~\cite{PhysRevA.100.063803,qj87-5xz9}.

More recently, topological band theory has provided a new framework for understanding BICs, which leads to the creation of topological BICs~\cite{PhysRevLett.113.257401,PhysRevLett.118.166803,PhysRevLett.132.046601}. These BICs are boundary-localized and protected by bulk topological invariants, and therefore exhibit extremely high robustness against disorders~\cite{meng2022merging,yin2024valley}. Such advances extend the scope of bulk-boundary correspondence from gapped systems to the continuum and further bridge the fields of topological physics and open wave systems~\cite{PhysRevB.101.161116,PhysRevLett.125.213901}. Meanwhile, higher-order topological phases has revealed the possibility of realizing lower-dimensional boundary states beyond conventional edges and surfaces, such as corner and hinge modes~\cite{benalcazar2017quantized,schindler2018higherorder,PhysRevLett.122.236401,PhysRevLett.123.216803,xie2021higherorder,PhysRevLett.130.116103,PhysRevB.108.L201102,lin2024probing,k4gb-14cg,qy19-f9wm}. These developments naturally motivate the exploration of higher-order topological BICs with corner localization inside the continuum~\cite{ni2019observation,PhysRevB.100.075120,hu2021nonlinear,wang2021quantum,PhysRevLett.130.106301,Zhang2024,PhysRevB.111.075427}. Despite many significant progresses in both topological BICs and higher-order topological phases, the corner-localized topological BICs in three-dimensional (3D) systems remain unexplored. How to engineer a controllable number of corner-localized topological BIC also becomes an interesting fundamental issue. 

\begin{figure*}[t]
\centering
\includegraphics[width=2.0\columnwidth,trim=0.65cm 0cm 0cm 0cm,clip=false]{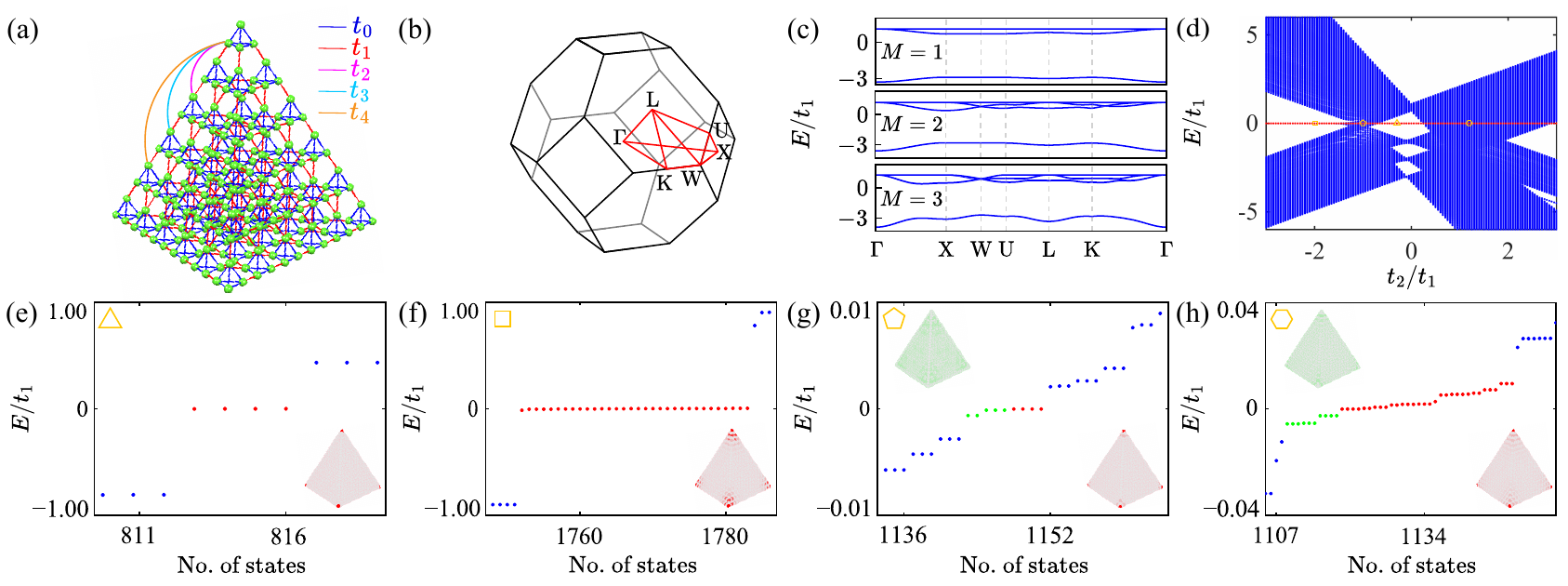}
\caption{(a) Schematic diagram of generalized 3D breathing pyrochlore lattice when taking $M=4$. (b) The Brillouin zone of Hamiltonian~\eqref{eq:Hamiltonian}, where $\Gamma=(0,0,0)$, $\text{X}=(2\pi,0,0)$, $\text{W}=(2\pi,0,\pi)$, $\text{U}=(2\pi,{\pi}/2,{\pi}/2)$, $\text{L}=({\pi},{\pi},{\pi})$ and $\text{K}=({3\pi}/2,0,{3\pi}/2)$ are six high symmetry points. (c) Band structures along $\Gamma$-$\text{X}$-$\text{W}$-$\text{U}$-$\text{L}$-$\text{K}$-$\Gamma$ line for $M=1$ with $t_0=0.1t_1$, $M=2$ with $(t_0/t_1,t_2/t_1)=(0.1,0.1)$, and $M=3$ with $(t_0/t_1,t_2/t_1,t_3/t_1)=(0.1,0.1,0.1)$. (d)-(h) The $k_{x,y,z}$-OBC energy spectrum when taking $M=2$ and $t_0=0.1t_1$, where the system size is $15\times 15\times 15$. The insets give the real-space distributions of zero-energy states for (e) $t_2=-0.3t_1$, (f) $t_2=-2t_1$, (g) $t_2=-t_1$, and (h) $t_2=1.2t_1$.}
\label{fig:1}
\end{figure*}

In this Letter, we propose that the corner-localized topological BICs can emerge in a class of 3D generalized breathing pyrochlore lattices with general inter-cell hoppings. These BICs coexist with numerous bulk extended modes while remaining spatially localized at the lattice corners, and their number can be arbitrarily tuned by changing the inter-cell hopping parameters. To exactly characterize these exotic BICs, we introduce the concept of polarized topological charges, i.e., topological nodal points in the Brillouin zone (BZ) that are experimentally measurable, revealing that the formation and evolution of corner-localized BICs are governed by the behavior of these polarized topological charges. This result opens an insight for understanding such real-space localization phenomenon through the information in momentum space. Furthermore, three distinct types of topological phase transitions (TPTs) are uncovered in the corner-localized topological BICs, which correspond to different bulk-gap-closing mechanisms and are captured by the emergence of zero-polarization topological charges. Our work establishes a general framework for realizing and characterizing the controllable corner-localized topological BICs in 3D systems and provides a promising route toward their experimental observation. 

{\it\color{blue} Lattice and corner-localized topological BICs.}~We start from a class of generalized breathing pyrochlore lattice with general inter-cell hoppings, as shown in Fig.~{\ref{fig:1}}(a). Since each unit cell contains four lattice sites forming a tetrahedron in real space, the corresponding momentum-space Hamiltonian hosts four bands and reads
\begin{equation}
\begin{split}
H(\mathbf{k})=&-\left(
\begin{array}{cccc}
0 & h_{12} & h_{13} & h_{14} \\
h_{12}^{\ast} & 0 & h_{23} & h_{24} \\
h_{13}^{\ast} & h_{23}^{\ast} & 0 & h_{34}\\
h_{14}^{\ast} & h_{24}^{\ast} & h_{34}^{\ast} & 0%
\end{array}%
\right),
\end{split}
\label{eq:Hamiltonian}
\end{equation} 
in which these off diagonal elements are given by
$h_{ij}=t_0+\sum_{m}t_me^{-\mathrm{i}m\mathbf{k}\cdot\mathbf{a}_{ij}}$ with $m=1,2,\cdots, M$, where $t_0$ and $t_{m}$ denote the intra-cell and the inter-cell $m$th neighbor hopping rates, respectively. Here, $M$ denotes the maximum hopping between two unit cells that the system drives. The momentum is $\mathbf{k}=(k_x,k_y,k_z)$ and the BZ is a truncated octahedron in momentum space, as shown in Fig.~\ref{fig:1}(b). The lattice vectors are taken as $\mathbf{a}_{12}=(1/2,1/2,0)$, $\mathbf{a}_{13}=(0,1/2,1/2)$, $\mathbf{a}_{14}=(1/2,0,1/2)$, $\mathbf{a}_{23}=(-1/2,0,1/2)$, $\mathbf{a}_{24}=(0,-1/2,1/2)$, and $\mathbf{a}_{34}=(1/2,-1/2,0)$. 
Without the inter-cell long-range hoppings, i.e. $M=1$, we find that $H(\mathbf{k})$ describes a standard breathing pyrochlore lattice~\cite{PhysRevLett.120.026801}. Its band structures along $\Gamma$-$\text{X}$-$\text{W}$-$\text{U}$-$\text{L}$-$\text{K}$-$\Gamma$ line are shown in Fig.~\ref{fig:1}(c), where two degenerated flat bands emerge. However, the addition of inter-cell long-range hoppings removes the degeneracy of two flat bands and transforms them into the ordinary bands with nonzero dispersion, as shown in Fig.~\ref{fig:1}(c). Remarkably, there are always three bands that become a three-fold degenerate energy level at $\Gamma$ point, which are independent of the parameters of the system, as shown in Fig.~\ref{fig:1}(c). This nontrivial property implies that such lattice system only has one band gap and exhibits either an insulator or a metal, which is determined by the location of zero energy level. 

We uncover that the above generalized breathing pyrochlore lattice can host single or multiple zero-energy topological corner states, which are located not only in the bulk gap but also in the continuous. This distinction depends on that the former system is an insulator but the latter system is a metal. The maximum number of zero-energy states at each corner is $N_\text{max}=M^3$. These zero-energy corner states are topologically protected by time-reversal symmetry ${\Theta}{H}\left(\mathbf{k}\right){\Theta^{-1}} ={ H}\left(-\mathbf{k}\right)$ with $\Theta=\mathsf{K}$ and generalized chiral symmetry ${H}(\mathbf{k})+\Gamma_4{H}(\mathbf{k})\Gamma_4^{-1}+\Gamma_4^2{H}(\mathbf{k})\Gamma_4^{-2}+\Gamma_4^3{H}(\mathbf{k})\Gamma_4^{-3}=0$ with $\Gamma_4=\text{diag}(1,\mathrm{i},-1,-\mathrm{i})$, where $\mathsf{K}$ denotes a complex conjugation operator. We take $M=2$ to demonstrate these results in Fig.~\ref{fig:1}(d), where one and eight zero-energy states occur at each corner for $-t_1-t_0<t_2<t_1-t_0$ and the remaining parameter regions, respectively. The $k_{x,y,z}$-OBC energy spectra further confirm that these zero-energy states in the bulk gap are completely localized at four corners, identifying the system as a third-order topological insulator, as shown in Figs.~\ref{fig:1}(e) and \ref{fig:1}(f). However, the zero-energy states in the continuum occur not only at four corners but also extend into the bulk. Such states do not hybridize with the surrounding bulk states of the lattice even in the absence of a bulk gap, implying that these corner states are topological BICs, as shown in Figs.~\ref{fig:1}(g) and \ref{fig:1}(h). Unlike the traditional BICs by coupling in-gap corner states to the continuous spectrum~\cite{PhysRevB.101.161116,PhysRevLett.125.213901}, the emergence of these corner-localized topological BICs is attributed to the unique characteristic of this lattice when it holds a metallic phase.

{\it\color{blue}Emergence mechanism of corner-localized topological BICs.} To deeply understand the above topological properties, we develop a generic method to solve the 1D momentum-space edge Hamiltonians of this generalized breathing pyrochlore lattice in a tetrahedron-shaped geometry, revealing the emergent mechanism of corner-localized topological BICs. Firstly, we focus on any one of four corners in the lattice and redefine three new momenta $k_s$, where $s=\alpha,\beta,\gamma$ denote three different directions along $1$D real-space boundaries starting from the corner. For convenience, we hereby choose
\begin{equation}
k_\alpha=\mathbf{k}\cdot \mathbf{a}_{12},~~k_\beta=\mathbf{k}\cdot \mathbf{a}_{13},~~k_\gamma=\mathbf{k}\cdot \mathbf{a}_{14},
\end{equation}
which induce $\mathbf{k}\cdot \mathbf{a}_{23}=k_\beta-k_\alpha$, $\mathbf{k}\cdot \mathbf{a}_{24}=k_\gamma-k_\alpha$, and $\mathbf{k}\cdot \mathbf{a}_{34}=k_\gamma-k_\beta$. 
By taking the annihilation operators of the fermions on the $j$th unit cell as $c_j=(c_{1,j},c_{2,j},c_{3,j},c_{4,j})^\text{T}$, where $i=1,2,3,4$ are the indexes of four sites in a unit cell, we define the $m$th-order forward and $m$th-order backward shift operators as $\delta_s^m$ and $\delta_s^{*m}$ to drive $
\delta_s^m c_{i,j}=c_{i,j+ms}$ and $\delta_s^{*m}c_{i,j}=c_{i,j-ms}$, respectively~\cite{PhysRevResearch.2.043136}. These shift operators obey the properties of $\delta_s\delta^m_s=\delta^{m+1}_s$, $\delta^*_s\delta^m_s=\delta^{m-1}_s$, and $\delta^*_s=\delta^\dagger_s$ an infinite system. And then, all electron hoppings in this lattice can be described by these shift operators. We thus rewrite the momentum-space bulk Hamiltonian $H(\mathbf{k})$ as a second-quantized Hamiltonian $\tilde{H}=\sum^{+\infty}_{j=-\infty}c_j^\dagger\mathcal{H}c_j$ in an infinite real space. Here $\mathcal{H}$ can be regarded as a first-quantized Hamiltonian defined on the lattice and is given by $\mathcal{H}=\mathcal{V}+ \sum_{s,m}(\mathcal{K}_{s,m}\delta_s^{*m}+\mathcal{K}^\dagger_{s,m}\delta^m_s)/2$, with the hopping matrices being
\begin{equation}
\begin{split}
&~~~~~~~\mathcal{V}=-t_0
\begin{pmatrix}
		0 & 1 & 1 & 1\\[2pt]
		1 & 0 & 1 & 1\\[2pt]
		1 & 1 & 0 & 1\\[2pt]
		1 & 1 & 1 & 0
	\end{pmatrix},
	\mathcal{K}_{\alpha,m}=-t_m
\begin{pmatrix}
		0 & 2 & 0 & 0\\[2pt]
		0 & 0 & 0 & 0\\[2pt]
		0 & \delta_\beta^m  & 0 & 0\\[2pt]
		0 & \delta_\gamma^m & 0 & 0
	\end{pmatrix},\\
&\mathcal{K}_{\beta,m}=-t_m
\begin{pmatrix}
		0 & 0 & 2 & 0\\[2pt]
		0 & 0 & \delta_\alpha^m & 0\\[2pt]
		0 & 0 & 0 & 0\\[2pt]
		0 & 0 & \delta_\gamma^m & 0
	\end{pmatrix},
\mathcal{K}_{\gamma,m}=-t_m
\begin{pmatrix}
		0 & 0 & 0 & 2\\[2pt]
		0 & 0 & 0 & \delta_\alpha^m\\[2pt]
		0 & 0 & 0 & \delta_\beta^m\\[2pt]
		0 & 0 & 0 & 0
	\end{pmatrix}.
	\end{split}
\end{equation}
The details have been provided in Supplementary Material~\cite{SuppInfo}). Under this framework, the 1D edge Hamiltonians of generalized breathing pyrochlore lattice can be directly obtained, as we shall demonstrate in the following. Note that this result cannot be achieved through traditional edge theory~\cite{PhysRevLett.121.096803,PhysRevLett.123.177001}, although this theory has been successfully applied to understand numerous higher-order topological phases~\cite{PhysRevResearch.5.L022032,PhysRevB.107.045118}.

We next choose a set of reference states $\psi_{0,s}$. The existence of $s$-direction 1D edge states requires $\mathcal{K}_{s^\prime,m}\psi_{0,s}=0$, where  $s^\prime \in \{\alpha,\beta,\gamma\}$ and $s^\prime \neq s$. This restricts the forms of reference states to being $\psi_{0,\alpha}=(\chi_1,\chi_2,0,0)^\text{T}$, $\psi_{0,\beta}=(\chi_1,0,\chi_2,0)^\text{T}$, and $\psi_{0,
\gamma}=(\chi_1,0,0,\chi_2)^\text{T}$, where $\chi_{1}$ and $\chi_{2}$ are two functions. We then assume a Bloch-type wave function $
\Psi_s=\psi_{0,s}e^{\mathrm{i}(k_s j_s+\sum_{s^\prime \neq s}K_{s^\prime}j_{s^\prime})}$ with decaying exponentially to be the $s$-direction edge states, where $K_{s^\prime}$ is a complex wave number~\cite{SuppInfo}. By further solving $\mathcal{H}\Psi_s=\epsilon\Psi_s$, the 1D momentum-space edge Hamiltonian is finally given by
\begin{equation}
	{\cal{H}}_\text{e}(k_s)=-
	\begin{pmatrix}
		0& t_0+\sum\limits_{m}t_me^{-\mathrm{i}mk_s}  \\
		t_0+\sum\limits_{m}t_me^{\mathrm{i}mk_s} &0 \\
	\end{pmatrix},
	\label{NN_Pyrochlore_H_edge}
\end{equation}
where $k_s\in[0,2\pi)$ is an effective BZ. We observe that ${\cal{H}}_\text{e}(k_s)$ is only $k_s$-dependent and similar to the Bloch Hamiltonian of a $1$D Su–Schrieffer–Heeger (SSH) chain with general inter-cell hoppings. The similar edge Hamiltonian can also be obtained for the remaining three edges~\cite{SuppInfo}. It implies that six edges of the generalized breathing pyrochlore lattice in the tetrahedron-shaped geometry are the same generalized SSH chain, where each corner corresponds to a crossed endpoint of three generalized SSH chains. Since the maximum number of zero-energy states for each generalized SSH chain is $N^\text{SSH}_\text{max}=M$, the maximum number of zero-energy states of each corner is $N_\text{max}=M^3$. This key property determines the emergence of rich corner-localized topological BICs, when taking the reasonable parameters of the system, as shown in Figs.~\ref{fig:1}(d), \ref{fig:1}(g), and \ref{fig:1}(h).

\begin{figure}[!t]
	\centering	\includegraphics[width=1.0\columnwidth,trim=0.65cm 0cm 0cm 0cm,clip=false]{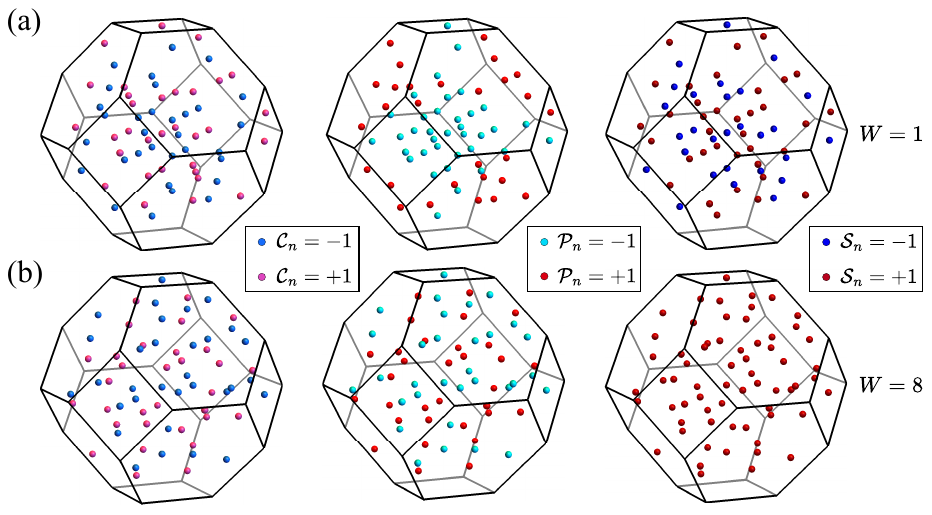}
\caption{Numerical results of topological charges, charge polarizations, and polarized topological charges for (a) $t_2=-t_1$ and (b) $t_2=1.2t_1$, giving $W=1$ and $W=8$, respectively. The other parameters are $t_0=0.1t_1$ and $M=2$.}
	\label{fig:2}
\end{figure}

{\it\color{blue}Momentum-space topological characterization.}~We next provide an exact topological characterization scheme for the zero-energy corner states through the polarized topological charges, which can be directly measured in realistic experiments~\cite{PhysRevB.110.L201117,rg5t-sx41}. Firstly, the topological number of the 1D edge Hamiltonian is proposed by rewriting $\mathcal{H}_{\text{e}}(k_s) = d_s'(k_s)\sigma_x + d_s(k_s)\sigma_y$ with
$d_s'(k_s)=-t_0-\sum_{m} t_m \cos (m k_s)$ and $d_s(k_s)=-\sum_{m} t_m \sin (m k_s)$, where $\sigma_{x,y,z}$ are three Pauli matrices. It allows $\mathcal{H}_\text{e}(k_s)$ to be characterized by a 1D winding number 
\begin{equation}
W_s= \frac{1}{2} \sum_{l_s} \mathrm{sgn}\left[{\partial_{k_s} d_s(\varrho_{l_s})}\right]\mathrm{sgn}[d_s^{\prime}(\varrho_{l_s})].
\end{equation}
Here $\varrho_{l_s}$ denotes the $l$th nodal point in the effective BZ $k_s$, which is determined by $d_s(\varrho_{l_s})=0$ with $l_s=1,2,\cdots,N_s$. Particularly, each nodal point can hold a topological charge ${\mathcal{C}_{l_s}^{k_s}}=\mathrm{sgn}[\partial_{k_s} d_s(\varrho_{l_s})]$ with the polarization ${\mathcal{P}_{l_s}^{k_s}}=\mathrm{sgn}[ d_s^{\prime}(\varrho_{l_s})]$. One can define a polarized topological charge as ${\mathcal{S}_{l_s}^{k_s}}={\mathcal{C}_{l_s}^{k_s}}{\mathcal{P}_{l_s}^{k_s}}$ to capture the topological properties of $\mathcal{H}_\text{e}(k_s)$ in momentum space.

Considering that the corner of this $3$D lattice is a crossed endpoint of three generalized SSH chains along $\alpha$, $\beta$, and $\gamma$ directions, the zero-energy states of each corner are naturally characterized by
\begin{equation}
	W = W_{\alpha} W_{\beta} W_{\gamma}=\frac{1}{8}\sum_n \mathcal{S}_n,
	\label{P_charge}
\end{equation}
where $\mathcal{S}_n=\mathcal{C}_n\mathcal{P}_n$
defines the $n$th polarized topological charge in the  BZ. The $n$th topological charge reads $\mathcal{C}_n ={\mathcal{C}_{l_\alpha}^{k_\alpha}}{\mathcal{C}_{l_\beta}^{k_\beta}}{\mathcal{C}_{l_\gamma}^{k_\gamma}}= \operatorname{sgn}[{\partial_{k_\alpha} d_\alpha(\boldsymbol{\varrho}_n)}
{\partial_{k_\beta} d_\beta(\boldsymbol{\varrho}_n)}{\partial_{k_\gamma} d_\gamma(\boldsymbol{\varrho}_n)}]$ and its  polarization is written as $
\mathcal{P}_n = {\mathcal{P}_{l_\alpha}^{k_\alpha}}{\mathcal{P}_{l_\beta}^{k_\beta}}{\mathcal{P}_{l_\gamma}^{k_\gamma}}=\operatorname{sgn}[d_\alpha^{\prime}(\boldsymbol{\varrho}_n)d_\beta^{\prime}(\boldsymbol{\varrho}_n)d_\gamma^{\prime}(\boldsymbol{\varrho}_n)]$. All these topological charges are located at the nodal points $\boldsymbol{\varrho}_n\equiv\{\mathbf{k}\in \text{BZ}|d_\alpha(\mathbf{k})=d_\beta(\mathbf{k})=d_\gamma(\mathbf{k})=0\}$
and hold a general form of $\boldsymbol{\varrho}_n=(\varrho_{l_\alpha},\varrho_{l_\beta},\varrho_{l_\gamma})$ with $n=1,2,\cdots, N_{\alpha}N_{\beta}N_{\gamma}$.
Therefore, the number of zero-energy states at each corner is identified by one-eighth of the total polarized topological charge in the original BZ, providing an insight for understanding the corner-localized topological BICs based on the momentum-space information.

We numerically show the topological charges, charge polarization, and polarized topological charges for $M=2$, as shown in Fig.~\ref{fig:2}. It is seen that the BZ contains a total of 64 nodal points hosting nontrivial topological charges. When the parameter is $t_2=-t_1$, there are 16 positive and 12 negative topological charges carrying the positive polarization, while the remaining 16 positive and 20 negative topological charges exhibit the negative polarization, as shown in Fig.~\ref{fig:2}(a). It induces 36 positive and 28 negative polarized topological charges and thus yields a non-zero topological number $W=1$, implying that the system hosts single zero-energy corner state at each corner, as shown in Fig.~\ref{fig:1}(g). After changing the parameter into $t_2=1.2t_1$, we observe that all positive (negative) topological charges hold the positive (negative) polarization, as shown in Fig.~\ref{fig:2}(b). It gives $W=8$, implying that the system can host 8 zero-energy corner state at each corner, as shown in Fig.~\ref{fig:1}(h). Clearly, this topological characterization scheme can intuitively and exactly capture the zero-energy corner states in the bulk gap or in the continuum.

\begin{figure}[t]
	\centering	\includegraphics[width=0.94\columnwidth,trim=0.65cm 0cm 0cm 0cm,clip=false]{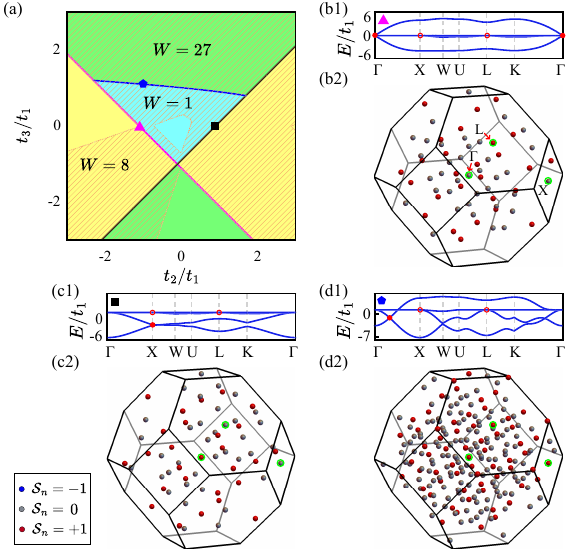}
	\caption{(a) Topological phase diagram determined by $W$, where the shaded regions mark the metallic phase. Three distinct phase boundaries correspond to the bulk gap closing at $\Gamma$ (pink), X (black), and certain non-high-symmetry point (blue), respectively. (b1)-(d1) The band structures with $(t_2/t_1,t_3/t_1)=(-1.1,0)$, $(0.9,0)$, and $(-1,1.1)$. (b2)-(d2) The distribution of polarized topological charges in the BZ. The green circles denote these polarized topological charges be located at the $\Gamma$, L and X points. The other parameters are $t_0=0.1t_1$ and $M=3$.}
	\label{fig:3}
\end{figure}

{\it\color{blue}Three types of TPTs in corner-localized topological BICs.}~We next demonstrate three distinct types of TPTs in the corner-localized topological BICs, which are induced by different bulk-gap-closing mechanisms and can be identified by the behavior of polarized topological charges. Based on Eq.~\eqref{P_charge}, we first numerically obtain phase diagram for $M=3$, as shown in Fig.~\ref{fig:3}(a). The corner-localized topological BICs emerge in shadow regions and are characterized by $W=1,8$, and $27$, giving three separate phase boundaries. Remarkably, these phase boundaries can also be theoretically obtained by solving $\sqrt{d_s^2+d_s^{\prime2}}=0$, i.e., $d_s=0$ and $d_s^{\prime}=0$~\cite{SuppInfo}. The explicit expressions are $t_3=-t_0-t_1-t_2$, $t_0-t_1+t_2$, and $(t_1+\sqrt{4t_0^2-4t_0t_2+t_1^2})/2$, which correspond to the pink, black, and blue lines, respectively. On these three phase boundaries, the energy gap of ${\cal{H}}_\text{e}(k_s)$ are clearly closed at $k_s=0, \pi$, and $\pm\text{arctan}\left[\frac{\sqrt{4t_1t_3+12t_3^2-2t_2(t_2+t_{\pm})}}{t_2+t_{\pm}}\right]$ with $t_{\pm}=\pm\sqrt{t_2^2+4t_3(-t_1+t_3)}$. Such nontrivial property leads to three different bulk-gap-closing mechanisms. Namely, type-I TPT: the bulk gap closes at $\Gamma$ point, as shown in Fig.~\ref{fig:3}(b1); type-II TPT: the bulk gap closes at X point, as shown in Fig.~\ref{fig:3}(c1); type-III TPT: the bulk gap closes at certain non-high-symmetry points, as shown in Fig.~\ref{fig:3}(d1). It is noted that two identical bands at X point and L point, which exhibit double degeneracy and are independent on the system parameters, are denoted by red circles. This feature determines that the remaining two bands should close energy gap in TPTs, marked by red solid points. When type-I, type-II, and type-III TPTs lead to a decrease in the number of corner-localized topological BICs, these BICs are naturally transformed into the mixture of bulk, surface, and hinge states, the individual bulk states, and the mixture of surface and hinge states, respectively. 

We further show that these TPTs can be identified by the emergence of zero-polarization topological charges and be distinguished by their different behaviors at high symmetry points, as shown in Figs.~\ref{fig:3}(b2), \ref{fig:3}(c2), and \ref{fig:3}(d2). For type-I TPT, the zero-polarization topological charge appear at $\Gamma$ and X points, which are associated with $k_s=0$, and thus there is a nonzero polarized topological charge at L point, as shown in Fig.~\ref{fig:3}(b2). For type-II TPT, the zero-polarization charge appear at L and X points, which are associated with $k_s=\pi$, and thus there is a nonzero polarized topological charge at $\Gamma$ point, as shown in Fig.~\ref{fig:3}(c2). For type-II TPT, the zero-polarization topological charge appear at certain adjustable momentum points, which are associated with $k_s=\pm\text{arctan}\left[\frac{\sqrt{4t_1t_3+12t_3^2-2t_2(t_2+t_{\pm})}}{t_2+t_{\pm}}\right]$, and thus there are three nonzero polarized topological charges at $\Gamma$, L, and X points, as shown in Fig.~\ref{fig:3}(d2). This means that the different TPTs of corner-localized topological BICs in real space can be faultlessly captured by the information of polarized topological charges. 

{\it\color{blue}Discussion and Conclusion.} For the Hamiltonian~\eqref{eq:Hamiltonian}, a standard breathing pyrochlore lattice is described by taking $M=1$, which has been realized in Li$\text{A}$Cr$_4$O$_8$ ($\text{A}$=In,Ga)~\cite{PhysRevLett.110.097203,PhysRevB.93.174402,PhysRevMaterials.9.033602}, Li$\text{A}$Cr$_4$S$_8$ ($\text{A}$=In,Ga)~\cite{PhysRevB.97.134117,okamoto2018magnetic}, and Ba$_3$Yb$_2$Zn$_5$O$_{11}$~\cite{PhysRevB.90.060414}. Apart from the electronic properties, such as 3D flat bands and 3D Dirac cones that are presented in this unique lattice~\cite{PhysRevLett.103.206805,wakefield2023three,huang2024observation,huang2024non}, it has attracted a lot of interest in the context of complex magnetism and geometric
frustration~\cite{li2016weyl,PhysRevLett.124.127203,PhysRevB.101.075118,PhysRevMaterials.7.104404,PhysRevB.109.064421,PhysRevB.105.235120}. Nevertheless, our work actually expands the understanding for the electronic properties of this lattice to the topological layer, particularly by developing a universal edge theory and revealing the corner-localized topological BICs to open an insight for the exploration of its broad range of novel physics. Moreover, although our 3D model focuses on a class of special long-range hoppings, such configurations are not limited to theoretical constructs and can be implemented in artificial systems, such as topological circuits~\cite{PhysRevLett.126.146802,wang2020circuit,yu20204d,10.1063/5.0265293,PhysRevB.98.201402} and photonic crystals~\cite{garcia2006band,PhysRevLett.100.013901,Ducrot18}. 

In summary, we have proposed a class of 3D generalized breathing pyrochlore lattice driving the controllable corner-localized topological BICs. We have developed an effective edge theory to profoundly reveal the physical reasons for the emergence of such topological BICs. Particularly, we have introduced polarized topological charges to exactly characterize the corner-localized topological BICs intertwining with a substantial number of bulk zero modes. It provides a new perspective for understanding the corner-localized topological BICs based on the momentum-space information. We have also uncovered three types of TPTs in the corner-localized topological BICs associated with different bulk-gap-closing mechanisms, which are intuitively identified by the emergence of zero-polarization topological charges at specific momenta. This work not only promotes the deeper study of corner-localized topological BICs but also provides a solid theoretical support for their future experimental observation.

{\it\color{blue}Acknowledgements.} This work is supported by the National Natural Science Foundation of China (Grant No. 12404318 and No. 12247101), the Fundamental Research Funds for the Central Universities (Grant No. lzujbky-2024-jdzx06), the Natural Science Foundation of Gansu Province (No. 22JR5RA389 and No. 25JRRA799), and the ``Talent Scientific Fund of Lanzhou University''.
\bibliography{myrefs}

\pagebreak
\clearpage
\onecolumngrid
\flushbottom
\begin{center}
\textbf{\large Supplementary Material of ``Topological bound states in the continuum with controllable multiplicity''}
\end{center}
%%%%%%%%%%%%%%%%%%%%%%%%%%%%%%%%%%%%%%%%%%%
\setcounter{equation}{0}
\setcounter{figure}{0}
\setcounter{table}{0}
\makeatletter
\renewcommand{\theequation}{S\arabic{equation}}
\renewcommand{\thefigure}{S\arabic{figure}}
\renewcommand{\bibnumfmt}[1]{[S#1]}
\renewcommand{\citenumfont}[1]{S#1}

In this Supplementary Material, we provide the details of edge theory and topological phase transitions (TPTs) in main text. In Sec.~I, we give a complete derivation of edge theory and obtain the 1D momentum-space edge Hamiltonians of generalized breathing pyrochlore lattice. In Sec.~II, we provide more numerical results about three types of TPTs in corner-localized topological bound states in the continuum (BICs).

\subsection{I. A complete derivation of edge theory}\label{one}
\subsubsection{a. Lattice Hamiltonian in an infinite real space}
Inspired by Ref.~\cite{s-PhysRevResearch.2.043136}, we define the forward and backward shift operators as $\delta f_j=f_{j+1}$ and $\delta^{*} f_j=f_{j-1}$. And then, the $m$th-order forward and $m$th-order backward shift operators are written as
\begin{equation}
\delta^m f_j=f_{j+m},~~\delta^{*m} f_j=f_{j-m},
\label{forward_backward_shift_operator}
\end{equation}
where $f_j$ is a function of the integer sequence $j$. For a system defined on the semi-infinite chain of lattice sites $j \geqslant 1$, we can show the following summation by parts:
\begin{equation}
\begin{aligned}
&\sum_{j=1}^{\infty}f_{j}\delta^m g_{j}=\sum_{j=m+1}^{\infty}(\delta^{*m}f_{j})g_{j}=\sum_{j=1}^{\infty}(\delta^{*m}f_{j})g_{j}-\sum_{j=1}^{m}f_{j-m}g_{j},\\ 
&\sum_{j=1}^{\infty}(\delta^m f_{j})g_{j}=\sum_{j=m+1}^{\infty}f_{j}\delta^{*m}g_{j}=\sum_{j=1}^{\infty}f_{j}\delta^{*m}g_{j}-\sum_{j=1}^{m}f_{j}g_{j-m},
\end{aligned}
\label{boundary_summation_by_parts}
\end{equation}
where $f_{j-m}$ and $g_{j-m}$ (with $j \leqslant m$) are assumed to be the states at the sites that outside the boundary of the chain. According to the properties of shift operators, we can use them to represent electron hopping in the lattice.

\begin{figure}[b]
\centering
\includegraphics[width=0.55\columnwidth,trim=0cm 0 0cm 0cm,clip=false]{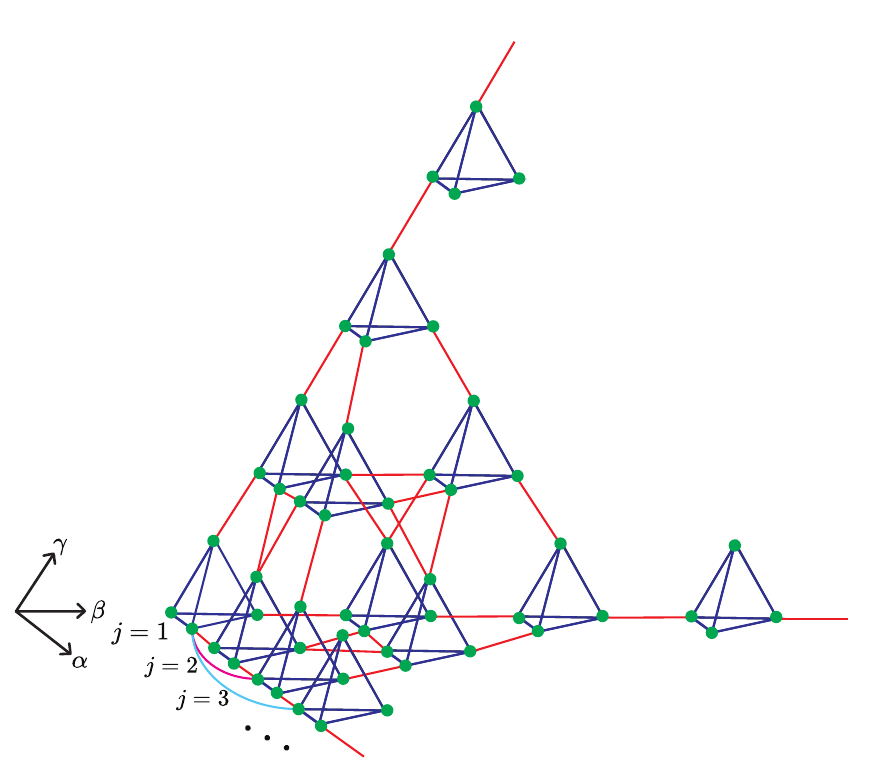}
\caption{Schematic of generalized breathing pyrochlore lattice with long-range hoppings. This lattice only shows the part with L=3, where $s \in {\alpha, \beta, \gamma}$ denotes three different directions along 1D real-space boundaries starting from one corner of the lattice. Here $j$ is the unit cell index along $s$ direction.
}
\label{fig:sm1}
\end{figure}

As shown in Fig.~{\ref{fig:sm1}}, we establish a basis-vector coordinate system for the generalized breathing pyrochlore lattice, where the directions of the basis vectors are denoted as $s = \alpha,\beta,\gamma$. We further define a unit cell as a small regular tetrahedron composed of four lattice sites. Then, each unit cell is marked by $j = (j_\alpha, j_\beta, j_\gamma)$ and the fermionic annihilation operators on the $j$th unit cell are denoted as $c_j = (c_{1,j}, c_{2,j}, c_{3,j}, c_{4,j})$. By using the identity  $\sum_{j=-\infty}^{\infty}f_{j+m}g_j=\sum_{j=-\infty}^{\infty}f_j g_{j-m}=\sum_{j=-\infty}^{\infty}f_j\delta^{*m} g_j$, the second-quantized Hamiltonianis for the bulk system can be written as
\begin{equation}
	\begin{aligned}
		H &= - \sum_{j=-\infty}^{\infty}\sum_{m=1}^{M}\Bigl[
		c_{1,j}^\dagger t_0 c_{2,j}
		+ c_{1,j}^\dagger t_0 c_{3,j}
		+ c_{1,j}^\dagger t_0 c_{4,j} + c_{2,j}^\dagger t_0 c_{3,j}
		+ c_{2,j}^\dagger t_0 c_{4,j}
		+ c_{3,j}^\dagger t_0 c_{4,j} + c_{1,j+m\alpha}^\dagger t_m c_{2,j}\\
		& + c_{1,j+m\beta}^\dagger t_m c_{3,j} + c_{1,j+m\gamma}^\dagger t_m c_{4,j} + c_{2,j+m(\beta-\alpha)}^\dagger t_m c_{3,j}
		+ c_{2,j+m(\gamma-\alpha)}^\dagger t_m c_{4,j}
		+ c_{3,j+m(\gamma-\beta)}^\dagger t_m c_{4,j}
		\Bigr] + \text{h.c.} \\
		&= - \sum_{j=-\infty}^{\infty}\sum_{m=1}^{M}\Bigl[
		c_{1,j}^\dagger (t_0 + t_m \delta_{\alpha}^{*m}) c_{2,j}
		+ c_{1,j}^\dagger  (t_0 + t_m \delta_{\beta}^{*m})c_{3,j} +c_{1,j}^\dagger (t_0 + t_m \delta_{\gamma}^{*m})  c_{4,j}
		+ c_{2,j}^\dagger (t_0 + t_m \delta_{\alpha}^{m} \delta_{\beta}^{*m}) c_{3,j} \\
		&  + c_{2,j}^\dagger (t_0 + t_m \delta_{\alpha}^{m} \delta_{\gamma}^{*m}) c_{4,j}+  c_{3,j}^\dagger(t_0 + t_m \delta_{\beta}^{m} \delta_{\gamma}^{*m}) c_{4,j}
		\Bigr] + \text{h.c.} \\
		&= \sum_{j=-\infty}^{\infty} c_j^{\dagger} \hat{\mathcal{H}} c_j
	\end{aligned}
	\label{second_quantized_Hamiltonian}
\end{equation}
where $\hat{\mathcal{H}}$ can be regarded as the first-quantized Hamiltonian and defined as
\begin{equation}
\hat{\mathcal{H}}=-\begin{pmatrix}
		0 & t_0 + \sum_m t_m \delta_{\alpha}^{*m} & t_0 + \sum_m t_m \delta_{\beta}^{*m} & t_0 + \sum_m t_m \delta_{\gamma}^{*m}\\
		t_0 + \sum_m t_m \delta_{\alpha}^m & 0 & t_0 + \sum_m t_m \delta_{\alpha}^m \delta_{\beta}^{*m} & t_0 + \sum_m t_m \delta_{\alpha}^m \delta_{\gamma}^{*m} \\
		t_0 + \sum_m t_m \delta_{\beta}^m & t_0 + \sum_m t_m \delta_{\alpha}^{*m} \delta_{\beta}^m & 0 & t_0 + \sum_m t_m \delta_{\beta}^m \delta_{\gamma}^{*m} \\
		t_0 + \sum_m t_m \delta_{\gamma}^m & t_0 + \sum_m t_m \delta_{\alpha}^{*m} \delta_{\gamma}^m & t_0 + \sum_m t_m \delta_{\beta}^{*m} \delta_{\gamma}^m & 0
	\end{pmatrix}.
	\label{first_quantized_Hamiltonian}
\end{equation}
Based on the above results, we write the intra‑cell hopping matrix $\mathcal{V}$ and the inter‑cell hopping matrices $\mathcal{K}_{s,m}$ as 
\begin{equation}
	\begin{array}{cc}
		\mathcal{V}=-t_0
		\begin{pmatrix} 
			0 & 1 & 1 & 1\\ 
			1 & 0 & 1 & 1 \\ 
			1 & 1 & 0 & 1 \\ 
			1 & 1 & 1 & 0 \\ 
		\end{pmatrix},
		\mathcal{K}_{\alpha,m}=-t_m
		\begin{pmatrix} 
			0 & 2 & 0 & 0 \\ 
			0 & 0 & 0 & 0 \\ 
			0 & \delta_{\beta}^m & 0 & 0 \\ 
			0 & \delta_{\gamma}^m & 0 & 0 \\ 
		\end{pmatrix}, 
		\mathcal{K}_{\beta,m}=-t_m
		\begin{pmatrix} 
			0 & 0 & 2 & 0 \\ 
			0 & 0 & \delta_{\alpha}^m & 0 \\ 
			0 & 0 & 0 & 0 \\ 
			0 & 0 & \delta_{\gamma}^m & 0 \\ 
		\end{pmatrix},
		\mathcal{K}_{\gamma,m}=-t_m
		\begin{pmatrix} 
			0 & 0 & 0 & 2 \\ 
			0 & 0 & 0 & \delta_{\alpha}^m \\ 
			0 & 0 & 0 & \delta_{\beta}^m \\ 
			0 & 0 & 0 & 0 \\ 
		\end{pmatrix}.
	\end{array}
	\label{V_K}
\end{equation}
Finally, the Hamiltonian $\hat{\mathcal{H}}$ can be compactly rewritten as
\begin{equation}
	\hat{\mathcal{H}}=\mathcal{V}+\frac{1}{2}\sum_{s,m}(\mathcal{K}_{s,m}\delta_s^{*m}+\mathcal{K}_{s,m}^{\dagger}\delta_s^m).
	\label{H}
\end{equation}
This is our key Hamiltonian, which has been provided in the main text.

\subsubsection{b. Boundary conditions, edge states, and 1D momentum-space edge Hamiltonians}

When the system has boundaries, the summation by parts for the shift operators produces additional boundary terms associated with each boundary. For each direction $s$, we introduce a boundary between $j_s=0$ and $j_s=1$. As a concrete illustration of the boundary effects, we consider the following two types of hopping terms on the semi‑infinite chain,
  \begin{equation}
 	\sum_{j=1}^{\infty}\sum_{m}c_{j+ms}^\dagger  c_{j}=\sum_{j=1}^{\infty}c_{j}^\dagger \delta_{s}^{*m} c_{j}-\sum_{j=1}^{m}c_{j}^\dagger c_{j-ms},~~ \sum_{j=1}^{\infty}\sum_{m}c_{j}^\dagger  c_{j+ms}=\sum_{j=1}^{\infty}c_{j}^\dagger \delta_{s}^{m} c_{j}.
 	\label{shift_half_infty}
 \end{equation}
The first type acquires a boundary correction, while the second does not. We write $H$ for the system defined on the semi‑infinite space $j_\alpha, j_\beta,j_\gamma \geqslant 1$ as
\begin{equation}
H= \sum_{j=1}^{\infty} c_j^{\dagger} \hat{\mathcal{H}}c_j-\frac{1}{2}\sum_{s,m}\sum_{j=1}^{m}c_j^{\dagger} \mathcal{K}_{s,m} c_{j-ms},
\label{Hamiltonian_boundary}
\end{equation}
where the operator $c_{j-ms}$ is outside the boundary in the $s$ direction. The boundary terms break the Hermiticity of the Hamiltonian, which must therefore be explicitly verified by appropriate boundary conditions. For any wave functions $\phi_j$ and $\psi_j$, we have 
\begin{equation}
\begin{aligned}
\langle\phi \mid\hat{\mathcal{H}} \psi\rangle 
&= \sum_{j=1}^{\infty} \phi_j^{\dagger} \hat{\mathcal{H}}\psi_j \\
&= \sum_{j=1}^{\infty} \phi_j^{\dagger} \bigl[\mathcal{V}+\frac{1}{2}\sum_{s,m}(\mathcal{K}_{s,m}\delta_s^{*m}+\mathcal{K}_{s,m}^{\dagger}\delta_s^m)\bigr]\psi_j \\
&= \sum_{j=1}^{\infty} \Bigl[ \phi_j^{\dagger}\mathcal{V}\psi_j+\frac{1}{2}\sum_{s,m}( \phi_j^{\dagger}\mathcal{K}_{s,m}\overleftarrow{\delta_s^{m}}\psi_j  +
\phi_j^{\dagger}\mathcal{K}_{s,m}^{\dagger}\overleftarrow{\delta_s^{*m}}\psi_j)\Bigr]-\frac{1}{2}\sum_{s,m}\sum_{j=1}^{m}(\phi_{j-ms}^{\dagger}\mathcal{K}_{s,m}^{\dagger}\psi_{j}-\phi_j^{\dagger}\mathcal{K}_{s,m}\psi_{j-ms})\\ 
&=\langle \hat{\mathcal{H}} \phi \mid\psi\rangle -\frac{1}{2}\sum_{s,m}\sum_{j=1}^{m}(\phi_{j-ms}^{\dagger}\mathcal{K}_{s,m}^{\dagger}\psi_{j}-\phi_j^{\dagger}\mathcal{K}_{s,m}\psi_{j-ms}),
\end{aligned}
\label{Hermiticity}
\end{equation}
where $\overleftarrow{\delta_s^{m}}$ and $\overleftarrow{\delta_s^{*m}}$ act on the left. Since the wave functions $\phi_j$ and $\psi_j$ are arbitrary for $j\ge 1$, we can derive the Hermiticity conditions
\begin{equation}
	\phi_{j-ms}^{\dagger}\mathcal{K}_{s,m}^{\dagger}\psi_j = 0, \quad
	\phi_j^{\dagger}\mathcal{K}_{s,m}\psi_{j-ms} = 0, \qquad \forall~s,m, j,
	\label{Hermiticity_condition}
\end{equation}
where $\psi_{j-ms}$ ($\phi_{j-ms}$) are assumed to be reference states outside the boundaries. Note that the following reference states chosen according to the boundary conditions must also satisfy this Hermiticity condition.

Next, we discuss the boundary conditions. Assume that the system is defined on the semi-infinite space $j_\beta \geqslant 1$ and $j_\gamma \geqslant 1$. Substituting Eq.~(\ref{H}) into the Schrödinger equation $\hat{\mathcal{H}} \psi_{jn} =\varepsilon_n \psi_{jn}$ yields
\begin{equation}
	\hat{\mathcal{H}}\psi_{j,\beta} =\mathcal{V}\psi_{j,\beta}+\frac{1}{2}\sum_{m}(\mathcal{K}_{\beta,m}\psi_{j-m\beta}+\mathcal{K}_{\beta,m}^{\dagger}\psi_{j+m\beta})=\varepsilon \psi_{j,\beta}~~(j=1,2,\dots),
	\label{beta_Seq}
\end{equation}
where the energy quantum number $n$ has been suppressed. For the boundary at $j_{\beta}=1$, the corresponding condition reads
\begin{equation}
	\mathcal{K}_{\beta,m}\psi_{j-m\beta}=0~~(m=1,2,3,\dots,M).
	\label{bc_beta}
\end{equation}
Likewise, at $j_\gamma=1$, one has
\begin{equation}
	\mathcal{K}_{\gamma,m}\psi_{j-m\gamma}=0~~(m=1,2,3,\dots,M).
	\label{bc_gamma}
\end{equation}
Consequently, when the system is defined on the $j_s \geqslant 1$, the boundary condition in the $s$ direction takes the form
\begin{equation}
	\mathcal{K}_{s,m}\psi_{j-ms}=0~~(m=1,2,3,\dots,M).
	\label{bc_s} 
\end{equation}
 From Eq.~(\ref{V_K}), we see that each matrix $\mathcal{K}_{s,m}$ contains only three nonzero elements, all located in the same column $c$ with $c=2,3,4$ for $s=\alpha,\beta,\gamma$, respectively. It follows that a reference state $\psi_0$ with a vanishing $c$th component satisfies $\mathcal{K}_{s,m}\psi_0=0$. Moreover, since the nonzero elements of $\mathcal{K}_{s,m}$ occupy the same matrix positions for all $m$, the boundary condition in Eq.~(\ref{bc_s}) is fulfilled by taking identical reference states,  
\begin{equation}
	\psi_0=\psi_{-1}=\psi_{-2}=\dots=\psi_{1-m}.
\end{equation}
Clearly, this reference state also respects the Hermiticity condition in Eq.~\eqref{Hermiticity_condition}. In conclusion, the Hermiticity, the lattice termination, and the boundary condition are all consistently by a suitable reference state $\psi_0$.

Finally, we study the hinge states propagating along the $\alpha$ direction, which arise from the intersection of boundaries in the $\beta$ and $\gamma$ directions. The Fourier transformation along the $\alpha$ direction is implemented by replacing $\delta_{\alpha}^m\rightarrow e^{imk_{\alpha}}$. According to the boundary conditions $\mathcal{K}_{\beta,m}\psi_{0n,\alpha}=0$ and $\mathcal{K}_{\gamma,m}\psi_{0n,\alpha}=0$, the reference state can be chosen as
\begin{equation}
	\psi_{0n,\alpha}= \begin{pmatrix} \chi_{1n} \\ \chi_{2n} \\ 0 \\ 0 \end{pmatrix},
\end{equation}
where $\chi_{1n}$ and $\chi_{2n}$ are nonzero components. We then assume a Bloch-type hinge state based on $\psi_0$
\begin{equation}
	\psi_{jn,\alpha} = \psi_{0n,\alpha} e^{iK_{n,\beta} j_\beta + iK_{n,\gamma} j_\gamma + i k_{\alpha} j_\alpha},
\end{equation}
with $j=(j_\alpha,j_\beta,j_\gamma)$, $K_{n,\beta}=k_{n,\beta}+i\kappa_{n,\beta}$, $K_{n,\gamma}=k_{n,\gamma}+i\kappa_{n,\gamma}$, and $\kappa_{n,\beta},\kappa_{n,\gamma} \geqslant 0$.  For brevity, we suppress the band index $n$ in the following. Substituting this ansatz into the eigenvalue equation yields
\begin{equation}
	-\sum_m \left(
	\begin{array}{cc|cc}		
		0 & t_0 + t_m e^{-im k_\alpha} & t_0 + t_m e^{-im K_{\beta}} & t_0 + t_m e^{-im K_{\gamma}} \\
		t_0 + t_m e^{im k_\alpha} & 0 & t_0 + t_m e^{im k_\alpha - im K_{\beta}} & t_0 + t_m e^{im k_\alpha - im K_{\gamma}} \\
		\hline
		t_0 + t_m e^{im K_{\beta}} & t_0 + t_m e^{-im k_\alpha + im K_{\beta}} & 0 & t_0 + t_m e^{im K_\beta - im K_{\gamma}} \\
		t_0 + t_m e^{im K_{\gamma}} & t_0 + t_m e^{-im k_\alpha + im K_{\gamma}} & t_0 + t_m e^{-im K_\beta + im K_{\gamma}} & 0
\end{array}
\right)
	\begin{pmatrix} \chi_1 \\ \chi_2  \\ \hline 0 \\ 0 \end{pmatrix}
	= \varepsilon \begin{pmatrix} \chi_1 \\ \chi_2 \\ \hline 0 \\ 0 \end{pmatrix}.
\end{equation}
Therefore, the effective edge Hamiltonian takes the form
\begin{equation}
\mathcal{H}_{\text{e}}(k_\alpha)=-\sum_m \begin{pmatrix}
		0 &  t_0 + t_m e^{-im k_\alpha}\\
		 t_0 + t_m e^{im k_\alpha} & 0
	\end{pmatrix}
\end{equation}

In a similar manner, the reference states for the hinge modes propagating along the $\beta$ and $\gamma$ directions can be determined from the corresponding boundary conditions. Specifically, imposing $\mathcal{K}_{\alpha,m}\psi_{0n,\beta}=0$ and $\mathcal{K}_{\gamma,m}\psi_{0n,\beta}=0$, as well as $\mathcal{K}_{\alpha,m}\psi_{0n,\gamma}=0$ and $\mathcal{K}_{\beta,m}\psi_{0n,\gamma}=0$, we obtain
\begin{equation}
	\psi_{0n,\beta}=
	\begin{pmatrix} 
		\chi_{1n}\\
		0\\
		\chi_{2n}\\
		0
	\end{pmatrix},
	\quad
	\psi_{0n,\gamma}=
	\begin{pmatrix} 
		\chi_{1n}\\
		0\\
		0\\
		\chi_{2n}
	\end{pmatrix},
\end{equation}
where $\chi_{1n}$ and $\chi_{2n}$ denote the nonvanishing components of the corresponding reference states. We then assume the Bloch-type wave functions
\begin{equation}
	\psi_{jn,\beta} = \psi_{0n,\beta}\, e^{iK_{n,\alpha} j_\alpha + iK_{n,\gamma} j_\gamma + i k_{\beta} j_\beta},
	\qquad
	\psi_{jn,\gamma} = \psi_{0n,\gamma}\, e^{iK_{n,\alpha} j_\alpha + iK_{n,\beta} j_\beta + i k_{\gamma} j_\gamma},
\end{equation}
with $K_{n,\alpha}=k_{n,\alpha}+i\kappa_{n,\alpha}$ and $\kappa_{n,\alpha}\geqslant 0$. For brevity, we suppress the band index $n$ in the following. Substituting these assumptions into the eigenvalue equation $\hat{\mathcal{H}}\psi_j=\varepsilon\psi_j$, we obtain the corresponding effective edge Hamiltonians,
\begin{equation}
	\begin{aligned}
		\mathcal{H}_{\mathrm{e}}(k_\beta) &=
		-\sum_m
		\begin{pmatrix}
			0 & t_0 + t_m e^{-im k_\beta}\\
			t_0 + t_m e^{im k_\beta} & 0
		\end{pmatrix},\\
		\mathcal{H}_{\mathrm{e}}(k_\gamma) &=
		-\sum_m
		\begin{pmatrix}
			0 & t_0 + t_m e^{-im k_\gamma}\\
			t_0 + t_m e^{im k_\gamma} & 0
		\end{pmatrix}.
	\end{aligned}
\end{equation}

\subsection{II. More numerical results of topological phase transitions}\label{two}

\begin{figure}[t]
	\centering	\includegraphics[width=1.0\columnwidth,trim=0cm 0 0cm 0cm,clip=false]{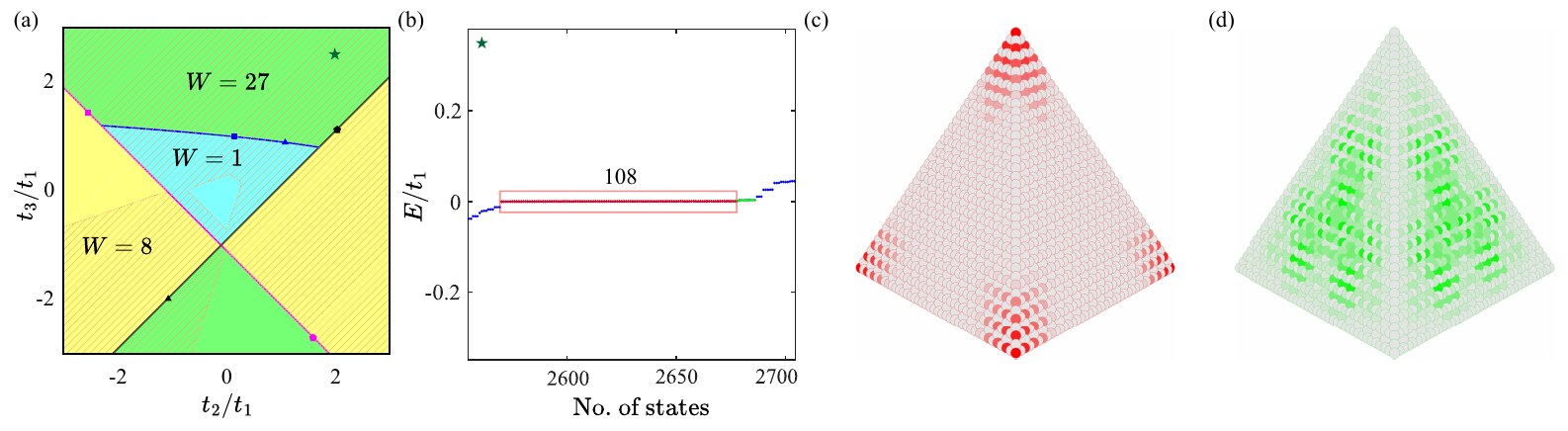}
	\caption{(a) Topological phase diagram determined by $W$. The orange shaded region marks the metallic phase. (b) Energy spectrum under open boundary conditions (OBC) at $(t_2/t_1,t_3/t_1)=(2,2.5)$ (green star). Red and green dots indicate zero-energy states (108 in red and 9 in green), and their real-space distributions are shown in (c) and (d), respectively. The number of unit cells along each edge is $L=20$. The other parameters are $t_0=0.1t_1$, $M=3$, and $t_{m>3}=0$.}
	\label{fig:sm2}
\end{figure}

Based on the above theory, the 1D momentum-space edge Hamiltonians of this generalized pyrochlore lattice are same as that of the generalized 1D Su–Schrieffer–Heeger (SSH) models. Accordingly, the zero-energy corner states at each corner, which are defined as the intersection of three generalized SSH chains along $\alpha$, $\beta$, and $\gamma$ directions, are naturally characterized by the product of $W_\alpha$, $W_\beta$, and $W_\gamma$, i.e.,  
\begin{equation}
W = W_{\alpha} W_{\beta} W_{\gamma}, 
\end{equation}
where $W_s$ with $s=\alpha,\beta,\gamma$ define a winding number of $\mathcal{H}_\text{e}(k_s)$~\cite{s-PhysRevB.110.L201117}. We take $M = 3$ as a example and obtain the topological phase diagram in Fig.~\ref{fig:sm2}(a). In metallic phase regions, all the zero-energy states can emerge both at the corners and in the bulk and/or surfaces, but these corner states do not hybridize with the bulk and/or surface states, leading to the corner-localized topological BICs. By choosing the parameters as $(t_2/t_1, t_3/t_1) = (2, 2.5)$, the corresponding $k_{x,y,z}$-OBC spectrum gives 217 zero-energy states, as shown in Fig.~\ref{fig:sm2}(b). Since the system host a nonzero topological number of $W = 27$, it means that each corner hosts 27 zero-energy states and yielding 108 corner-localized topological states, as shown in Fig.~\ref{fig:sm2}(c). The remaining 9 zero-energy states are naturally distributed on the surfaces, as shown in Fig.~\ref{fig:sm2}(d). 

\begin{figure}[h]
	\centering
	\includegraphics[width=0.9\columnwidth,trim=0cm 0 0cm 0cm,clip=false]{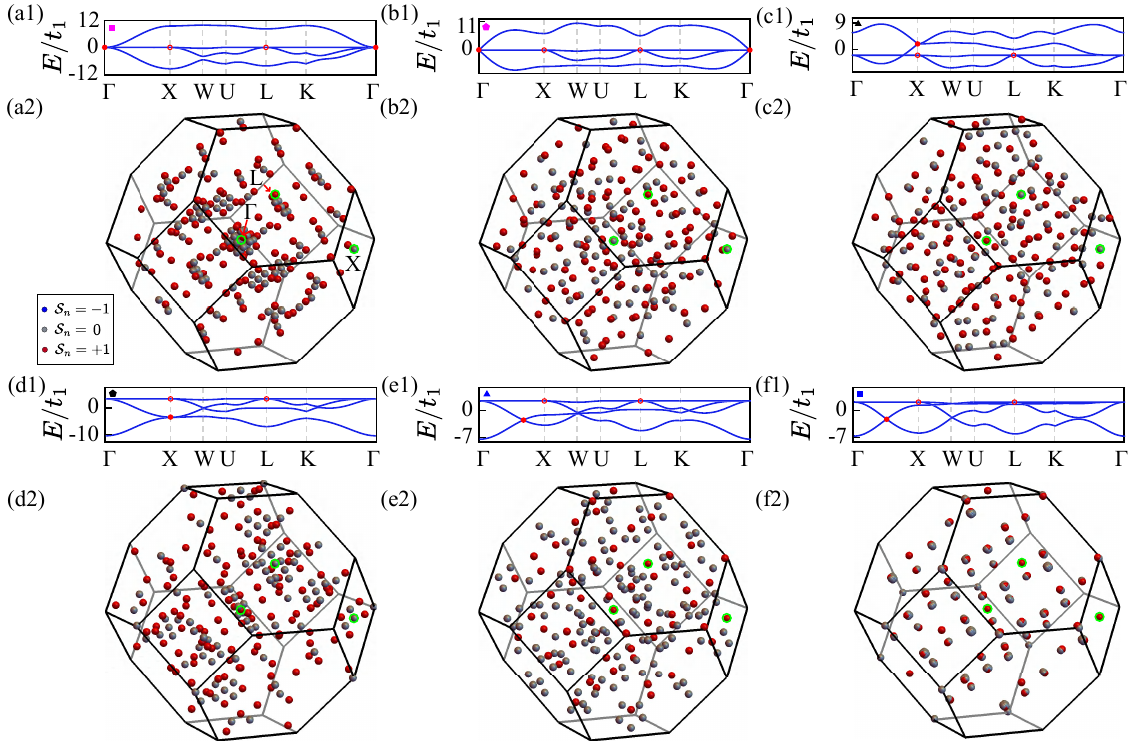}
	\caption{The bulk gap closes at high symmetry points on the phase boundaries for (a1) $(t_2/t_1,t_3/t_1)=(-2.5,1.4)$ (pink square), (b1) $(t_2/t_1,t_3/t_1)=(1.5,-2.6)$ (pink pentagon), (c1) $(t_2/t_1,t_3/t_1)=(-1,-1.9)$ (dark triangle), (d1) $(t_2/t_1,t_3/t_1)=(2,1.1)$ (dark pentagon), (e1) $(t_2/t_1,t_3/t_1)=(1,0.9)$ (blue triangle) and (f1) $(t_2/t_1,t_3/t_1)=(0.1,1)$ (blue square). (a2)-(f2) The corresponding distribution of polarized topological charges in the BZ. The green circles mark the topological charges located at the $\Gamma$, L and X points. The other parameters are $t_0=0.1t_1$, $M=3$,  and $t_{m>3}=0$.}
	\label{fig:sm3}
\end{figure}

Moreover, we observe three phase boundaries in the topological phase diagram, which are classified into three types based on their distinct bulk-gap-closing mechanisms. Namely, type-I TPT: the bulk gap closes at $\Gamma$ point, as shown in Figs.~\ref{fig:sm3}(a1) and \ref{fig:sm3}(b1); type-II TPT: the bulk gap closes at X point, as shown in Figs.~\ref{fig:sm3}(c1) and \ref{fig:sm3}(d1); type-III TPT: the bulk gap closes at certain non-high-symmetry points, as shown in Figs.~\ref{fig:sm3}(e1) and \ref{fig:sm3}(f1). By diagonalizing the 1D edge Hamiltonian $\mathcal{H}_e(k_s)$ to obtain the energy dispersion relation $\mathcal{E} = \pm E=\pm\sqrt{d_s^2 + {d_s^{\prime2}}}$ and the energy gap $\Delta=2E$, we derive the explicit expression of each phase boundary by solving $\Delta=0$, i.e.,  $d_s =0$ and $ d^{\prime}_s = 0$. Within the effective BZ $k_s \in [0, 2\pi)$, we seek all the solutions as follow: 
\begin{equation}
k_s=0,~\pi,~\pm\text{arctan}\left[\frac{\sqrt{4t_1t_3+12t_3^2-2t_2(t_2+t_{\pm})}}{t_2+t_{\pm}}\right],
\end{equation}
with $t_{\pm}=\pm\sqrt{t_2^2+4t_3(-t_1+t_3)}$. And then, we substitute these solutions into $d_s^{\prime} = 0$, and thus three phase boundaries are given by
\begin{equation}
t_3 = -t_0 - t_1 - t_2,~~t_0 - t_1 + t_2,~~\frac{t_1 + \sqrt{4t_0^2 - 4t_0 t_2 + t_1^2}}{2},
\end{equation}
which correspond to pink, black, and blue lines in Fig.~\ref{fig:sm2}(a), respectively. Since $d_s = 0$ determines the positions of topological charges $\boldsymbol{\varrho}_n \equiv \{\mathbf{k} \in \text{BZ} \mid d_\alpha(\mathbf{k}) = d_\beta(\mathbf{k}) = d_\gamma(\mathbf{k}) = 0\}$, but $d_s^{\prime} = 0$ identifies the vanishing polarization of these topological charges, i.e., $\mathcal{P}_n = \operatorname{sgn}[d_\alpha^{\prime}(\boldsymbol{\varrho}_n) d_\beta^{\prime}(\boldsymbol{\varrho}_n) d_\gamma^{\prime}(\boldsymbol{\varrho}_n)] = 0$, the zero-polarization  topological charges must emerge at $\mathbf{k}=(k_x,k_y,k_z)$ with
\begin{equation}
k_x=k_\alpha + k_\beta - k_\gamma,~~k_y=k_\alpha - k_\beta + k_\gamma,~~k_z=-k_\alpha + k_\beta + k_\gamma,
\end{equation} 
where $k_\alpha, k_\beta, k_\gamma \in \{0, \pi, \pm \arctan[{\sqrt{4t_1 t_3 + 12 t_3^2 - 2t_2(t_2 + t_{\pm})}}/{(t_2 + t_{\pm}})]\}$. Since $k_s=0$ and $\pi$ for $\Delta=0$ does not depends on the parameters $t_m$, their combinations yield eight momentum points in the 3D BZ with zero-polarization topological charges for $(0,0,0)$ (i.e., $\Gamma$ point), $(0,2\pi,0)$, $(2\pi,0,0)$ (i.e., X point), and $(0,0,2\pi)$ for type-I TPTs and $(\pi,-\pi,\pi)$, $(-\pi,\pi,\pi)$, $(\pi,\pi,-\pi)$, and $(\pi,\pi,\pi)$ (i.e., L point) for type-II TPTs, respectively. For type-III, it is clear that there is no zero-polarization topological charges to be located at $\Gamma$, X, and L points. These conclusions can be confirmed by the numerical results in Fig.~\ref{fig:sm3}. 


It is worth mentioning that we can intuitively understand the process of TPTs via the behavior of polarized topological charges. For example, when considering the TPT between $W=8$ and $27$, there are initially 91 zero-polarization topological charges and 125 positive polarized topological charges in the BZ, as shown in Figs.~\ref{fig:sm3}(a2), \ref{fig:sm3}(b2), \ref{fig:sm3}(c2) and \ref{fig:sm3}(d2). When the TPT occur and $W$ is increased, all 91 zero-polarization charges become positive polarized topological charges. This implies that the total number of positive polarized topological charges is 216, giving a topological number of $W=27$. When the TPT occur and $W$ is decreased, we observe that 15 of them become positive polarized topological charges and the remaining 76 become negative polarized topological charges. This means that there are 140 positive and 76 negative polarized topological charges, yielding a topological number of $W=8$. Similarly, for the TPT between $W=1$ and $27$, the BZ contains 152 zero-polarization topological charges and 64 positive polarized topological charges, as shown in Figs.~\ref{fig:sm3}(e2) and \ref{fig:sm3}(f2). When the TPT occur and $W$ is increased, all 152 zero-polarization topological charges become the positive values, and thus we obtain 216 positive polarized topological charges, which gives $W=27$. When the TPT occur and $W$ is decreased, 48 and 104 zero-polarization topological charges become positive and negative values, respectively. Then, we obtain 112 positive and 104 negative polarized topological charges, giving a topological number of $W=1$. Clearly, the TPTs show a conversion from zero-polarization to nonzero-polarization topological charges, providing a picture to understand the TPTs based on the information of momentum space.

\end{document}